\documentclass[10pt,conference]{IEEEtran}
\usepackage[utf8]{inputenc}
\usepackage[T1]{fontenc}
\usepackage{float}
\usepackage{amsmath}
\usepackage{amsfonts}
\usepackage{mathtools}
\usepackage{hyperref}
\IEEEoverridecommandlockouts
\usepackage{amssymb}
\usepackage{graphicx}
\usepackage{makeidx}
\usepackage{csquotes}
\usepackage{cite}
\usepackage{xcolor}
\usepackage{subcaption}
\usepackage{balance}
\usepackage{array,makecell,booktabs}
\newcolumntype{P}[1]{>{\raggedright\arraybackslash}p{#1}}
\newcolumntype{C}[1]{>{\centering\arraybackslash}p{#1}}
\usepackage{booktabs}
\usepackage{multirow}
\usepackage{array}
\usepackage{makecell}
\usepackage{tabularx}

\begin{document}

\title{Transmissive RIS-Assisted Vehicular Direct-to-Satellite Communications: Opportunities, Limitations, and Comparison with Phased Arrays}
\author{Wali Ullah Khan, Abdullah Abdullah, Zakir Ullah, Juan Andres Vasquez-Peralvo, Eva Lagunas

\thanks{Wali Ullah Khan, Abdullah Abdullah, Juan Andres Vazquez-Peralvo, and Eva Lagunas are with Interdisciplinary Centre for Security, Reliability and Trust (SnT), University of Luxembourg, Luxembourg (emails: \{waliullah.khan, abdullah.abdullah, juan.vasquez, eva.lagunas\}@uni.lu).}

\thanks{Zakir Ullah is with Centre for Wireless Communications-Radio Engineering (CWC-RE), University of Oulu, Finland (email: Zakir.Ullah@oulu.fi).}

\thanks{This research was funded by the Luxembourg National 
Research Fund (FNR) under the project VARRAY-5G 
(C24/IS/18968070). }
}%
\maketitle
\begin{abstract}
This article studies transmissive reconfigurable intelligent surface (RIS)-assisted architectures and compares them with electronically steered phased arrays for the deployment of vehicular direct-to-satellite (D2S) communications in future satellite networks. Rather than treating RIS as a direct replacement for phased arrays, we clarify the operating regimes in which RIS can serve as a low-power wavefront-shaping aperture and those in which phased arrays remain preferable because of their high gain and mature beam-tracking capability. Moreover, phased arrays can support multi-beam operation, which is particularly beneficial for dual connectivity and seamless handover. We distinguish analog, digital, and hybrid phased arrays, discuss the relationship between transmissive RIS and reconfigurable transmitarrays, and highlight practical profile, tracking, and link-budget constraints for mobile terminals. The comparison shows that passive RIS offers attractive power efficiency and aperture scalability, active RIS can partially improve the link budget, and phased arrays remain preferable for high‑throughput, fast‑tracking, multi‑beam links, although their cost is still prohibitive for large‑scale mass‑market deployment. Finally, open challenges and hybrid RIS-array design directions are discussed for future vehicular D2S systems.
\end{abstract}


\IEEEpeerreviewmaketitle

\section{Introduction}

Direct-to-satellite (D2S) connectivity is emerging as an important component of future sixth-generation (6G) and non-terrestrial networks, particularly for connected transportation, emergency response, intelligent logistics, and broadband access in poorly served regions. Vehicular terminals, however, must sustain directional links under mobility, blockage, and rapidly changing satellite geometry while operating within strict size, weight, power, thermal, and aerodynamic constraints. The beamforming architecture is therefore a central design choice rather than a secondary implementation detail.

Phased arrays remain the reference solution for satellite terminals because they provide high aperture gain, fast electronic steering, mature tracking, and, when supported by sufficient RF chains, flexible multi-beam operation. Their power and complexity must nevertheless be interpreted according to architecture. Analog arrays may realize efficient single-beam steering with one or a few RF chains, whereas highly parallel hybrid and fully digital arrays require substantially more converters, amplifiers, control electronics, and baseband processing. A fair comparison should therefore avoid attributing the power burden of digital beamforming to all phased-array implementations.

Reconfigurable intelligent surfaces (RISs) offer a complementary route based on programmable wavefront shaping with fewer RF chains \cite{10740042}. For vehicular D2S terminals, transmissive RIS is particularly relevant because an illuminated aperture can shape the outgoing field while separating the feed from the radiating direction \cite{10886969}. In practice, such a structure is closely related to a reconfigurable transmitarray and inherits feed-to-aperture spacing, illumination taper, spillover, insertion loss, and profile constraints. These issues are decisive for vehicle rooftop deployment, where a deep feed-illuminated terminal may be less attractive than a mature low-profile array.

Passive RIS offers low control power and scalable aperture size, but it cannot amplify the incident signal and is therefore limited in severe satellite link budgets. On the other hand, active RIS can amplify the incident signal at the expense of local noise \cite{10584518}. However, these two architectures of RIS cannot offer the full functionality of a digital phased array, having an independent data stream and simultaneous beams. Therefore, RIS can be viewed as a low-power aperture or front-end layer to reduce the complexity of a phased array.

In this paper, our objective is to present a balanced comparison between RIS and phased arrays for vehicular D2S communication. In particular, this comparison provides beamforming gain, power consumption, hardware complexity, scalability, tracking capability, terminal profile, and prototype maturity. Moreover, this study also distinguishes transmissive RIS from electronically steered phased-array replacement and identifies the operating regimes in which RIS can provide useful coverage shaping or link-budget assistance. In addition, this work discusses prototype design, qualitative normalized design maps, and open research directions for the practical deployment of future LEO vehicular satellite terminals.

\begin{figure*}[!t]
\centering
\includegraphics[width=\textwidth]{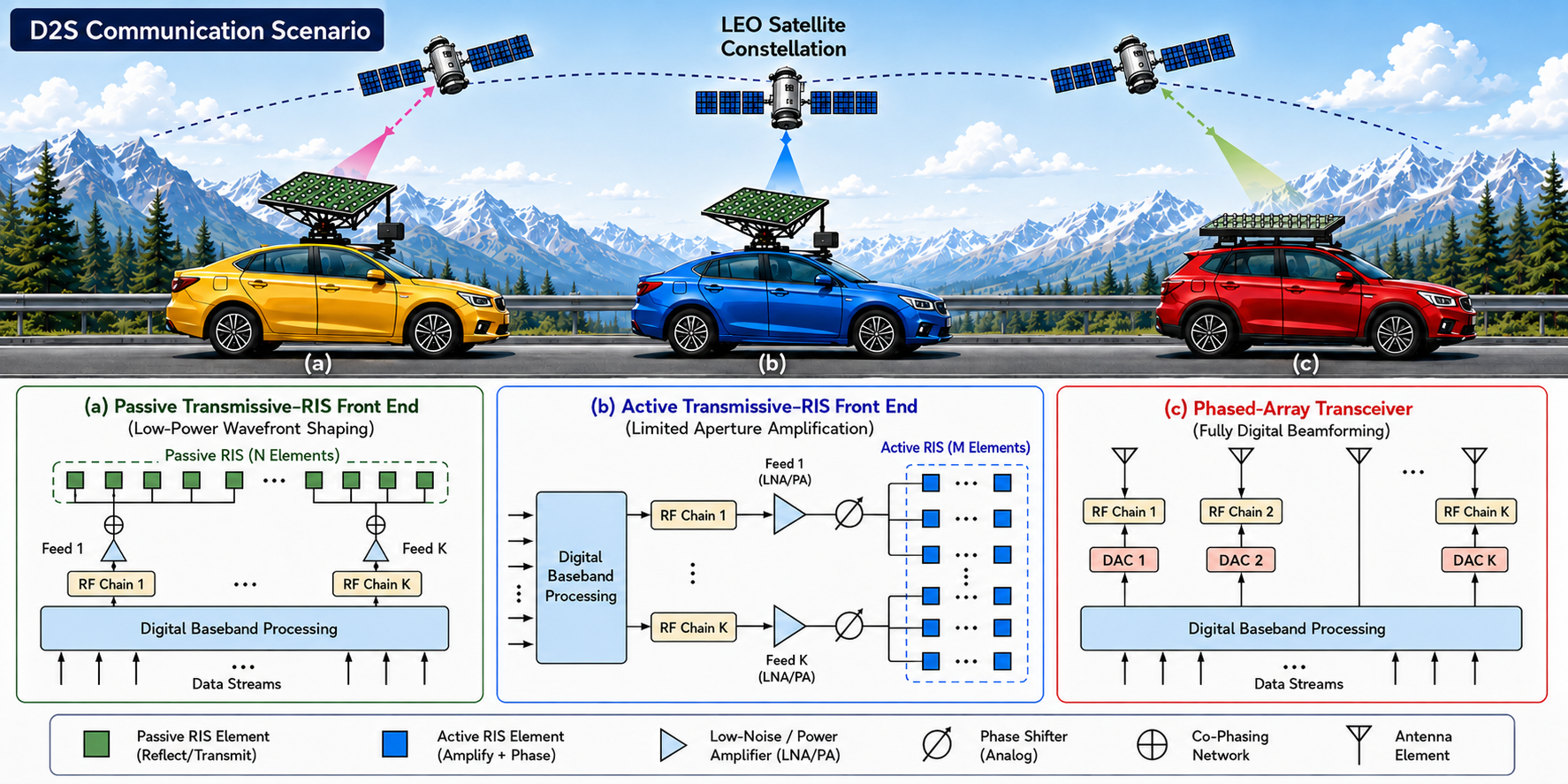}
\caption{Representative vehicular transceiver architectures for D2S communication: (a) a passive transmissive-RIS front end for low-power wavefront shaping, (b) an active transmissive-RIS front end with limited aperture amplification, and (c) a phased-array transceiver with fully digital beamforming.}
\label{fig:T_RIS}
\end{figure*}

\section{Beamforming Architectures for Vehicular Direct-to-Satellite Communications}
\label{sec:architectures}

Vehicular D2S terminals must balance link budget, steering speed, aperture size, power consumption, and terminal profile. LEO links emphasize continuous tracking and rapid handover, whereas MEO and GEO links place greater weight on propagation range and received-power margin. Fig.~\ref{fig:T_RIS} illustrates the three architecture families considered in this work: passive RIS, active RIS, and phased arrays.

\subsection{RIS-Based Transceiver Architectures}

A transmissive RIS is a programmable aperture whose unit cells control the phase, amplitude, or polarization of an incident electromagnetic field \cite{10689360}. When placed in front of a feed antenna, it can generate directive or shaped beams while requiring fewer RF chains than a fully active digital array. The practical suitability of transmissive RIS depends not only on performance but also on feed integration, aperture efficiency, mechanical depth, control complexity, and aerodynamic packaging.

\textbf{Transmissive RIS, transmitarrays, and profile constraints:}
A feed-illuminated transmissive RIS is closely related to a reconfigurable transmitarray. In both structures, an incident wave generated by a feed antenna passes through a planar aperture whose unit cells impose programmable phase and/or amplitude responses. Consequently, the terminology should be interpreted carefully because a practical transmissive RIS may inherit several transmitarray-related constraints, including feed-to-aperture spacing, illumination taper, spillover loss, insertion loss, blockage from supporting structures, and overall mechanical profile \cite{10839492}. Current vehicular roof‑mounted flat phased‑array/SATCOM terminals typically occupy a footprint of approximately 40–60 cm × 30–40 cm, with a profile height of about 4–8 cm above the roofline, as reported in commercial Ku/Ka‑band product datasheets (e.g., flat Starlink‑type and Kymeta u8 terminals). A large feed-to-aperture distance can make the overall terminal substantially deeper than a commercial low-profile phased-array terminal. Transmissive RIS architectures are therefore most attractive when combined with compact feeds, integrated feed--lens structures, folded electromagnetic paths, conformal metasurfaces, or hybrid RIS--array configurations that reduce the terminal profile.

\begin{table*}[t]
\centering
\caption{Comparison of phased-array beamforming architectures, highlighting the key differences among analog, hybrid, and fully digital implementations.}
\label{tab:beamforming_architectures}
\renewcommand{\arraystretch}{1.15}
\setlength{\tabcolsep}{5pt}
\begin{tabular}{p{3.3cm} p{3.2cm} p{3.2cm} p{3.2cm}}
\hline
\textbf{Parameter} &
\textbf{Analog} &
\textbf{Hybrid} &
\textbf{Fully Digital} \\
\hline
Beamforming domain
& RF phase shifters
& RF and digital domains
& Digital domain only \\

RF chains
& One
& Subarray-based
& One per antenna element \\

Simultaneous beams
& One
& 2--8, typically
& Many \\

Hardware complexity
& Low
& Medium
& Very high \\

ADC/DAC count
& Very low
& Moderate
& Very high \\

Calibration
& Simple
& Moderate
& Complex \\

Power consumption
& Lowest
& Medium
& Highest \\

Cost
& Lowest
& Medium
& Highest \\

Beam flexibility
& Low
& Medium
& Highest \\

Adaptive nulling
& No
& Limited
& Yes \\

Typical applications
& User terminals
& SATCOM terminals
& Base stations and radar systems \\
\hline
\end{tabular}
\end{table*}

\textbf{Passive and active RIS implementations:}
Passive RIS architectures use low-power tuning components and obtain aperture gain through coherent wavefront transformation rather than RF amplification. They are therefore attractive for size-, weight-, and power-constrained large apertures. Nevertheless, because a passive RIS cannot restore or amplify the incident signal power, insertion loss and feed loss can significantly limit its uplink and long-range performance. Active RIS architectures incorporate amplification, active matching, or gain-controllable components within the aperture. These capabilities can compensate for part of the propagation and insertion losses and improve the available link margin, particularly at high carrier frequencies. However, active RIS also introduces amplifier noise, nonlinear distortion, biasing networks, increased power consumption, thermal-management requirements, and additional reliability concerns. It therefore occupies an intermediate position between passive wavefront-shaping surfaces and conventional fully active antenna arrays.

\textbf{Diagonal and beyond-diagonal RIS control:}
RIS architectures can be further classified according to the degree of coupling and control among their unit cells. For example, each element in a conventional diagonal RIS independently controls the phase of the incident signal, which results in a diagonal starting matrix. This RIS architecture is simple to design and implement, but it offers limited control over the incident signal. In contrast, beyond-diagonal RIS provides controllable coupling among different elements through reconfigurable impedance or interconnection networks, resulting in strong scattering of the incident signal in the desired direction. This architecture provides joint control of signal phase and amplitude, thereby providing finer wavefront manipulation, improved interference suppression, multi-lobe radiation-pattern synthesis, and enhanced spatial-multiplexing capability. These features are potentially valuable for spectrum sharing, multi-user D2S access, and interference-limited satellite networks. Nevertheless, beyond-diagonal control also increases circuit complexity, control overhead, insertion loss, calibration requirements, and hardware-design difficulty, which must be carefully considered for vehicular and spaceborne deployments. It is important to note that RIS-based transceivers should not be viewed as direct one-to-one replacements for fully digital phased arrays. A passive transmissive RIS primarily transforms the wavefront generated by one or more feed antennas. Although appropriately designed surface configurations can produce shaped or multi-lobe radiation patterns, this does not imply that an RIS with $N$ elements can generate $N$ independently modulated simultaneous beams. RIS is therefore better interpreted as a low-power programmable aperture or wavefront-shaping layer that can complement a feed antenna, subarray, or compact phased-array front-end.

\begin{table*}[!t]
\centering
\caption{Representative phased-array prototypes for vehicular and satellite communications.}
\label{tab:phased_array_comparison}
\begin{tabular}{|p{3cm}|p{5cm}|p{4cm}|p{4cm}|}
\hline
\textbf{Parameter} & \textbf{Phased Array \cite{10071999}} & \textbf{Phased Array \cite{9803256}} & \textbf{Phased Array \cite{Hou2025PhasedAW}} \\
\hline
Fabrication & Multilayer PCB technology & Multilayer PCB technology & Multilayer PCB technology \\
\hline
Hardware complexity \& cost & High & Low & Low \\
\hline
Power consumption & 1.8 V, 5.7 A & 15 V & Not explicitly reported \\
\hline
Operational architecture & DC power, 8$\times$8 beamforming board, IDC cable, voltage level translator, interface controller, USB, computer & Beamforming chip, 16:1 Wilkinson power divider, DC power supply, FPGA board & 16 beamforming chips, LNAs, power divider, DC power supply, control system \\
\hline
Achievable gain & 29.7 dBi & 30.5 dBi & 26 dBi \\
\hline
Number of radiating elements & 64 & 64 & 256 \\
\hline
Impedance bandwidth & 22.3--43.4 GHz & 17.7--21.2 GHz & 17.2--21.2 GHz \\
\hline
Scan loss & 3 dB & 3.5 dB & $<$3 dB \\
\hline
Side-lobe level (SLL) & 13.3 dB & N/A & $<$15 dB \\
\hline
G/T & $-6.4$ dB/K & $-2.94$ dB/K & 0.21 dB/K (RHCP broadside) \\
\hline
Noise temperature & 303 K (load) & 50 K (antenna) & 80 K (antenna) \\
\hline
Physical aperture size & 8$\times$8 & 8$\times$8 & 16$\times$16 \\
\hline
Polarization & Dual CP & Dual CP & Dual CP \\
\hline
Scanning range & $\pm 53^\circ$ & $\pm 50^\circ$ & $\pm 45^\circ$ \\
\hline
Typical application scenarios & CubeSat communications & LEO SATCOM applications & LEO satellite communication \\
\hline
\end{tabular}%
\end{table*}

\subsection{Electronically Steered Phased-Array Architectures}
Electronically steered phased arrays are increasingly used in satellite user terminals and in some specialized vehicles (e.g., Starlink‑equipped platforms). However, their bulk and cost still limit large‑scale integration into commercial passenger cars, motivating research on compact, low‑cost solutions. Depending on the location and extent of signal processing, phased arrays can be classified as analog, hybrid, or fully digital architectures.

Analog phased arrays offer comparatively simple and low-power beam steering using RF phase shifters and a small number of RF chains. However, they generally support limited simultaneous-beam operation and suffer from constraints such as beam squint, restricted adaptive nulling, and reduced signal-processing flexibility. Fully digital arrays provide the highest beamforming flexibility, multi-beam capability, interference suppression, and advanced signal processing because each antenna element is connected to an independent RF chain and data converter. This flexibility comes at the expense of substantially higher hardware complexity, power consumption, calibration burden, thermal load, and cost.

Hybrid beamforming provides a compromise between these two architectures by combining analog beamforming in the RF domain with digital processing in the baseband domain \cite{lankshear2026hybrid}. A typical hybrid architecture performs:
\begin{itemize}
    \item analog beam steering within individual antenna subarrays using RF phase control; and
    \item digital beamforming across the outputs of the subarrays using a reduced number of RF chains.
\end{itemize}

Table~\ref{tab:beamforming_architectures} compares analog, hybrid, and fully digital phased-array architectures. This distinction is essential when comparing phased arrays with RIS-based terminals. Comparing a passive transmissive RIS directly with a fully digital array can be misleading because the latter provides independently controllable RF chains, simultaneous data streams, and digital multi-beam processing, whereas the former primarily performs passive or semi-active wavefront transformation.

Despite their technological maturity and strong link performance, phased arrays continue to present implementation challenges. Fully digital and highly parallel hybrid arrays require multiple RF chains, beamforming integrated circuits, feed networks, power dividers, amplifiers, high-speed ADCs and DACs, control circuits, calibration systems, and cooling mechanisms. These requirements become particularly restrictive in vehicular terminals, where rooftop area, available electrical power, aerodynamic profile, heat dissipation, and manufacturing cost are tightly constrained. Thus, although phased arrays currently provide the highest technological maturity and link-level capability, their scalability and cost-effectiveness remain important concerns for mass-market vehicular D2S deployment, particularly when simultaneous digital multi-beam functionality is required.

Table~\ref{tab:phased_array_comparison} summarizes representative phased-array prototypes reported in the literature, with emphasis on their antenna architectures and performance. The designs in \cite{10071999,9803256,Hou2025PhasedAW} employ analog phased-array architectures using a single RF input/output port together with beamforming integrated circuits for electronic steering. In \cite{10071999}, a multilayer $8\times8$ dual-circularly polarized array is developed for CubeSat communications, whereas \cite{9803256,Hou2025PhasedAW} consider phased-array designs for LEO satellite applications.

These prototypes demonstrate high antenna gain, broad electronic scanning ranges, and suitability for satellite terminals based on performance indicators such as gain-to-noise-temperature ratio, scan loss, and receiver noise temperature. However, they also highlight the implementation burden associated with beamforming integrated circuits, RF front ends, low-noise amplifiers, power dividers, control boards, calibration circuits, and DC power supplies. Such analog arrays can also serve as subarray building blocks for more advanced hybrid or fully digital beamforming architectures.
```

\subsection{Key Design Insight}

The main difference between passive RIS, active RIS, and phased arrays lies in how they trade performance for power and hardware complexity. Passive RIS provides the best energy efficiency and scalability, but its lack of amplification limits the achievable link-budget improvement. Active RIS improves received signal strength and offers a practical middle ground, but requires additional power, thermal management, and noise-aware design. Different antenna architectures, including passive RIS, active RIS, phased arrays and hybrid RIS–array solutions, may coexist depending on system requirements such as link budget, mobility, frequency of operation, terminal size and power constraints. Therefore, future vehicular D2S systems are unlikely to rely on a single universal architecture. 

\section{Prototype-Level Comparison of RIS and Phased Arrays}

Prototype-level comparison is needed because gain alone does not determine suitability for a vehicle-mounted satellite terminal. Power, cooling, aperture, steering range, implementation complexity, and scalability are equally important. Analog phased arrays steer directive beams with limited RF-chain count; digital arrays add independent data-conversion paths and flexible multi-beam processing; passive RIS shapes an incident field without per-element RF amplification; and active RIS adds limited gain while avoiding full RF-chain duplication. The following comparison therefore emphasizes engineering trade-offs rather than a single performance metric.

\subsection{Prototype-Level Comparison of RIS Architectures}

Table~\ref{tab:prototype_comparison_ris} summarizes representative passive RIS and active RIS prototypes reported in the recent literature. The selected works were chosen because they provide sufficient experimental details regarding aperture size, gain, power consumption, phase resolution, and scanning performance.

\begin{table*}[t]
\centering
\caption{Prototype-level comparison of passive RIS and active RIS from the selected literature.}
\label{tab:prototype_comparison_ris}
\renewcommand{\arraystretch}{1.15}
\setlength{\tabcolsep}{3pt}
\scriptsize
\resizebox{\textwidth}{!}{%
\begin{tabular}{|p{2.6cm}|p{2.3cm}|p{2.3cm}|p{2.3cm}|p{2.3cm}|p{2.5cm}|p{2.5cm}|}
\hline
\textbf{Parameters} & \textbf{Passive RIS \cite{9020088}} & \textbf{Passive RIS \cite{10177872}} & \textbf{Passive RIS \cite{diaby20192}} & \textbf{Active RIS \cite{10738302}} & \textbf{Active RIS \cite{rao2023active}} & \textbf{Active RIS \cite{wu2022wideband}} \\
\hline

Fabrication &
Multilayer PCB technology &
Multilayer PCB technology &
Multilayer PCB technology &
Multilayer PCB technology &
Multilayer PCB technology &
Multilayer PCB technology \\
\hline

Hardware complexity \& cost &
Intermediate &
Low &
Intermediate &
Low &
Moderate--High (PA + phase shifter per element) &
Low \\
\hline

Power consumption &
8.7 W (PINs) &
4.3 W (PINs) &
4.7 W (PINs) &
120 mA, 5.5 V ($P = 0.66$ W) &
DC bias $\approx 1.75$ V per element (current not explicitly reported); estimated order of 10s of watts for 16 elements &
12 V DC supply; current not reported \\
\hline

Operational architecture &
FPGA, control board, RIS board, Tx and Rx horn, signal generator, USRP, computer with required cables &
FPGA, control board, RIS board, Tx and Rx horn, signal generator, USRP with LabVIEW, computer with required cables &
FPGA, control board, RIS board, Tx and Rx horn, signal generator &
-- &
Reflection-type active RIS with phase-reconfigurable reflection amplifiers &
RIS embedded with PA; 2$\times$2 subarray power-combining architecture \\
\hline

Achievable gain &
21.7 dBi at 2.3 GHz and 19.1 dBi at 28.5 GHz &
22 dBi at 27 GHz &
19.8 dBi &
11.9 dB &
8.5 dB active gain per element &
12.2 dBi \\
\hline

Phase resolution &
2-bit (4 phases) &
2-bit (4 phases) &
2-bit &
2-bit (4 phases) &
2-bit (4 states) &
N/A \\
\hline

Number of radiating elements &
256 &
256 &
196 &
32 &
16 (4$\times$4 prototype) &
16 \\
\hline

Reflective / transmissive bandwidth &
2--2.6 GHz, N/A at 28 GHz &
25--28.5 GHz (estimated) &
26.2--30.9 GHz &
2.57--2.62 GHz &
2.2--2.6 GHz &
5--6 GHz \\
\hline

Scan loss &
3.7 dB &
4 dB (X) and 5 dB (Y) &
5 dB &
3.5 dB &
N/A &
N/A \\
\hline

Side-lobe level (SLL) &
16.7 dB &
22 dB &
N/A &
11 dB (estimated) &
N/A &
N/A \\
\hline

Physical aperture size &
16$\times$16 &
16$\times$16 &
14$\times$14 &
4$\times$8 &
4$\times$4 &
4 (2$\times$2 subarray) \\
\hline

Polarization &
Dual (LH and RH) &
Dual CP &
Single &
Single &
Single &
Single \\
\hline

Number of PINs &
1024 &
512 &
784 &
-- &
0 (uses transistors and RF switches instead) &
0 \\
\hline

Scanning range &
$\pm 60^\circ$ &
$\pm 60^\circ$ &
$\pm 60^\circ$ &
$\pm 60^\circ$ &
N/A &
$\pm 10^\circ$ \\
\hline

Typical application scenarios &
Mobile and satellite communications &
Satellite communication &
Satellite communications, electronically steerable transmitarrays &
Mobile communication &
Active RIS-aided wireless links and coverage extension &
Signal enhancement and coverage extension \\
\hline

\end{tabular}%
}
\end{table*}

The results in Table~\ref{tab:prototype_comparison_ris} reveal several important trends. First, passive RIS prototypes generally require only a few watts of DC power, since their operation relies mainly on PIN diodes or other low-power tuning elements. Even for large apertures containing several hundred unit cells, the total power consumption typically remains below 10~W. For example, the passive RIS prototype in \cite{9020088} consumes approximately 8.7~W despite containing 1024 PIN diodes, while the Ka-band transmitarray in \cite{diaby20192} requires approximately 4.7~W.

Second, active RIS architectures increase the received signal strength by incorporating distributed amplification inside the RIS aperture. This enables improved link performance relative to passive RIS, particularly in long-range satellite links where the free-space attenuation is severe. However, the addition of active components also increases the hardware complexity, thermal burden, and power demand. Depending on the architecture, the total DC consumption of active RIS can range from less than 1~W to several tens of watts.

It should be noted that the reported gain values for RIS are not directly comparable to the EIRP of phased array transmitters. Passive RIS does not amplify the RF signal or increase transmit power; instead, it enhances the received signal through coherent redirection and constructive combination of incident waves. Thus, the reported gain represents an improvement in the effective channel response rather than RF amplification. In contrast, active RIS provides limited signal amplification, while phased arrays increase EIRP through dedicated RF power amplifiers.

\subsection{Implications for Vehicular Direct-to-Satellite Links}
For vehicular D2S communications, the key design challenge is achieving high received signal strength, wide beam steering capability, and low power consumption simultaneously. For vehicular D2S communications, the key design challenge is to simultaneously achieve high received signal strength, with analog phased-array gains of approximately $26$--$30~\mathrm{dBi}$, a wide elevation beam-steering range, typically from about $-45^{\circ}$ to $+50^{\circ}$, and low terminal power consumption, ranging from approximately $10~\mathrm{W}$ to several tens of watts in reported analog implementations.
Passive RIS is particularly attractive for vehicular platforms because it can provide wide-angle coverage shaping with extremely low power consumption and a potentially conformal aperture. The prototypes considered in Table~\ref{tab:prototype_comparison_ris} demonstrate scanning ranges up to $\pm 60^\circ$ while maintaining power consumption below 10~W. Such characteristics are useful for LEO satellite visibility and pointing-error tolerance, although the lower peak gain of wider beams must be balanced against the link-budget requirements of a moving vehicle.

Passive RIS also suffers from limited link improvement because it cannot compensate for the severe path loss encountered in satellite channels. In contrast, active RIS provides moderate amplification, enabling stronger received signals while maintaining lower complexity than phased arrays. However, it offers more limited beam steering capability and requires additional thermal management. Conventional phased arrays remain the benchmark solution due to their mature high-gain beam tracking and reliable link performance. Their power consumption and hardware complexity depend on the adopted architecture: analog arrays can be relatively efficient, whereas fully digital and highly parallel hybrid arrays become more demanding due to their RF-chain and processing requirements.


\begin{table*}[!t]
\centering
\caption{Qualitative normalized design map for vehicular direct-to-satellite communications.}
\label{tab:normalized_comparison}
\begin{tabular}{|p{3.3cm}|p{3.4cm}|p{3.4cm}|p{3.8cm}|}
\hline
\textbf{Parameter} & \textbf{Phased Arrays} & \textbf{Passive RIS} & \textbf{Active RIS} \\
\hline
Operating frequency band & S / Ku / Ka (17--43 GHz reported) & S-band and Ka-band (2--30 GHz demonstrated) & S / C-band (2--6 GHz demonstrated) \\
\hline
Effective aperture size & Medium--large (8$\times$8 to 16$\times$16 elements) & Large (up to 16$\times$16, 14$\times$14 apertures) & Medium (4$\times$4 to 4$\times$8 prototypes) \\
\hline
Achievable gain & High (26--30.5 dBi measured) & Medium (19--22 dBi measured) & $\sim$8--12 dB active amplification reported in selected prototypes;
total aperture gain is implementation-dependent \\
\hline
Beam steering range & Wide ($\pm 45^\circ$ to $\pm 53^\circ$) & Wide ($\pm 60^\circ$) & Limited--moderate ($\pm 10^\circ$ to $\pm 30^\circ$ effective) \\
\hline
Power consumption & Medium–high (worst‑case digital) & Low ($\approx$4--10 W reported) & Moderate ($\approx$10--50 W estimated/reported) \\
\hline
Hardware complexity & Low--very high depending on RF-chain count and beamforming type & Low--intermediate (unit cells + bias/control) & Intermediate--high (active components, bias, control) \\
\hline
Thermal management requirement & Low--critical depending on RF power and digital/hybrid hardware & Negligible--low & Moderate \\
\hline
Scalability with aperture size & Architecture-dependent; RF-chain and PA count may limit scaling & Excellent (low-cost aperture scaling) & Good but limited by active-element count \\
\hline
Beam re-pointing/control requirement & High for narrow high-gain LEO tracking; lower for wider analog beams & Moderate; wider beams may reduce update frequency & Moderate--high depending on beamwidth and stability \\
\hline
Form factor / conformality & Mature low-profile panels exist; complexity grows with active hardware & Flat aperture, but feed-illuminated t-RIS may require depth & Good, but thermal/biasing hardware must be integrated \\
\hline
Uplink suitability & Excellent & Limited (double path loss) & Good (path-loss compensated) \\
\hline
Downlink suitability & Excellent & Excellent & Excellent \\
\hline
Vehicular D2S suitability & High for robust high-throughput links; cost/power depend on architecture & Complementary for low-power coverage shaping and downlink assistance & Complementary for moderate link-budget enhancement \\
\hline
\end{tabular}
\end{table*}

\section{System-Level Design Insights for Vehicular D2S}
\label{sec:normalized_comparison}

Tables~\ref{tab:phased_array_comparison} and~\ref{tab:prototype_comparison_ris} summarize representative phased-array and RIS prototypes. A direct numerical comparison among these technologies is nevertheless difficult because the implementations under consideration vary in terms of operating frequency, aperture size, feed architecture, number of RF chains or active elements, measurement configuration, and link assumptions, a direct numerical comparison between the reported prototypes is not strictly possible. Therefore, rather than being a like-for-like link-budget comparison, the analysis that follows is meant to be a \emph{technology-level comparative design map}. Rather than establishing an absolute performance ranking, the numerical values provided in the literature are used to highlight representative operating regions and architectural trade-offs. All three architectures would need to be assessed under a standard D2S scenario with the same assumptions regarding satellite orbit, carrier frequency, aperture area, bandwidth, transmit power/EIRP, receiver G/T, pointing loss, propagation loss, and implementation losses in order to conduct a thorough quantitative comparison.


Table~\ref{tab:normalized_comparison} therefore positions the three architectures according to their main system-level trade-offs. Phased arrays provide the highest beamforming flexibility, particularly when hybrid or digital processing supports multiple independent beams, although their power, thermal burden, and hardware complexity depend strongly on the selected architecture. Passive RIS offers excellent control-power efficiency and aperture scalability, but its lack of RF amplification restricts its ability to compensate for severe satellite path loss. 
By boosting the link budget without demanding the whole RF-chain duplication of a digital phased array, Active RIS holds a middle position. Therefore, rather than being proof that one design always performs better than the others, the comparison should be seen as an engineering technology map.

\begin{figure*}[!t]
    \centering
    \subfloat[]{
        \includegraphics[width=0.31\textwidth]{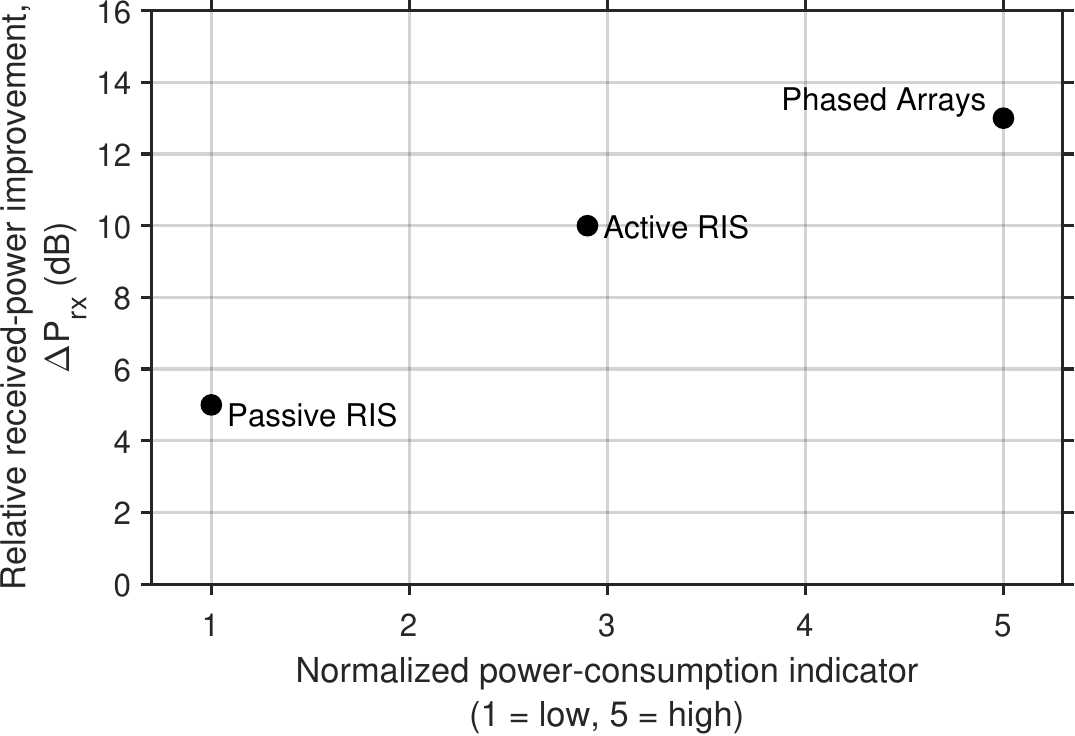}
        \label{fig:rxpower_vs_power}
    }
    \hfill
    \subfloat[]{
        \includegraphics[width=0.31\textwidth]{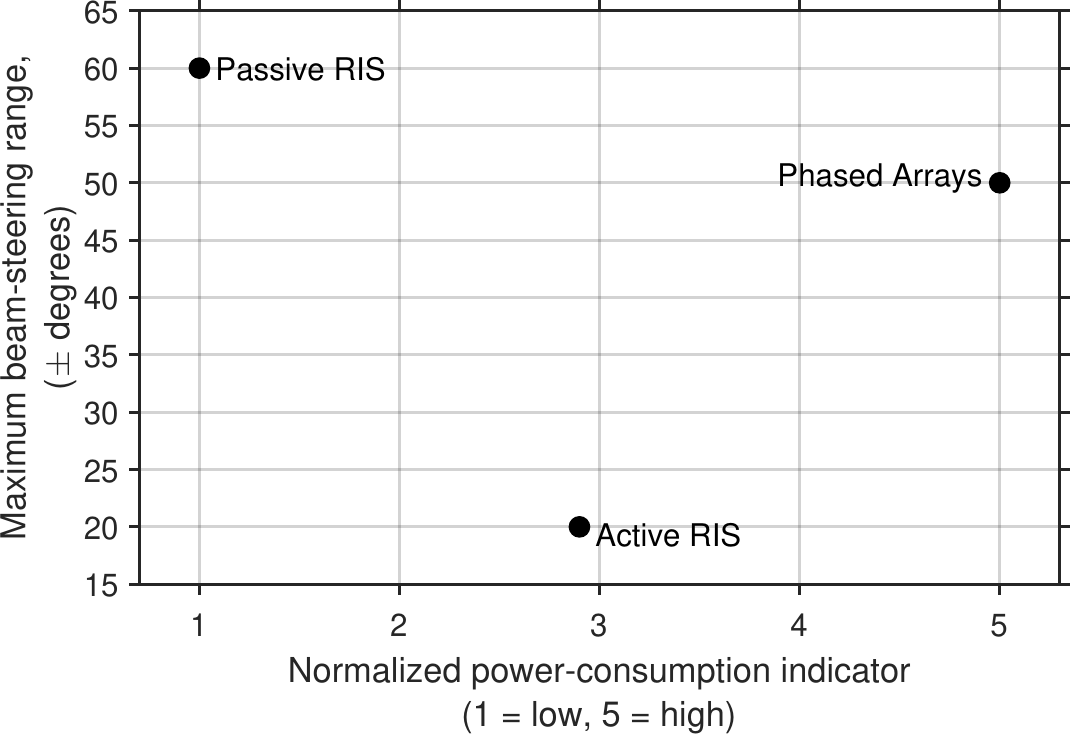}
        \label{fig:steering_vs_power}
    }
    \hfill
    \subfloat[]{
        \includegraphics[width=0.31\textwidth]{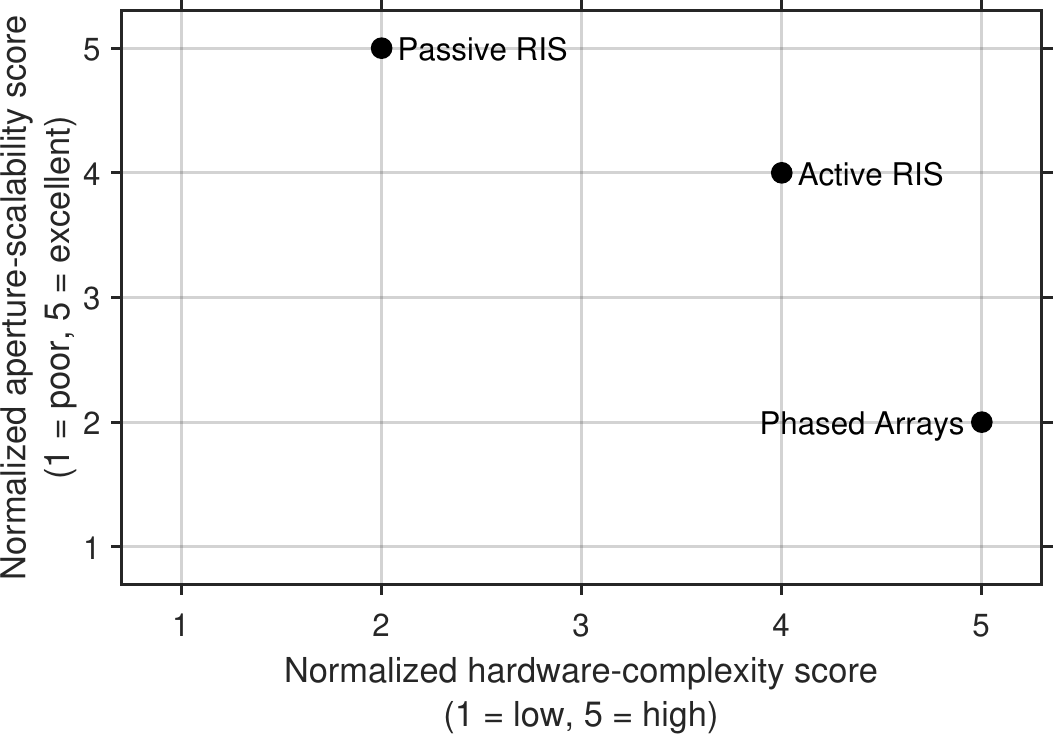}
        \label{fig:scalability_vs_complexity}
    }
    \caption{Qualitative normalized design map of passive RIS, active RIS, and phased-array technologies for vehicular D2S communications: (a) representative relative received-power improvement vs. normalized power-consumption indicator; (b) beam-steering range vs. normalized power-consumption indicator; and (c) normalized aperture-scalability score vs. normalized hardware-complexity score. In (a) and (b), lower x-axis values indicate lower relative power demand. In (c), the scores are qualitative ordinal ratings. The points summarize broad trends from heterogeneous prototype reports and are not intended as a direct same-satellite link-budget or hardware-level comparison.}
    \label{fig:system_level_comparison}
\end{figure*}

\subsection{Link-Budget Improvement and Power Consumption}

Fig.~\ref{fig:system_level_comparison}(a) shows a qualitative relationship between relative received-power improvement to the normalized power-consumption indicator. Due to differences in operating frequency, aperture size, hardware design, and testing settings between reported RIS and phased-array prototypes, the normalized axis has been used. Therefore, rather than directly comparing absolute power usage or a shared satellite connection budget, the displayed points show broad trends at the technical level.

Since passive RIS primarily modifies the incident wave without active RF amplification, it emerges at the low-power end. This makes it appealing for vehicle terminals with limited power and heat, but when the incident satellite signal is weak, the achievable received-power enhancement is still restricted. Active RIS shifts toward a higher power-consumption area by using active components to give more signal amplification. Although phased arrays provide great link improvement and high beamforming gain, their RF chains, amplifiers, and related electronics have the highest relative power requirements.

Thus, the trade-off that results depends on the application. While active RIS offers more link improvement at the expense of higher power and hardware overhead, passive RIS is appealing for low-power wavefront control. When the terminal requirements are dominated by maximal beamforming gain and connection dependability, phased arrays are still preferred. Therefore, active RIS may be thought of as a middle ground between completely active phased-array designs and passive aperture control.

\subsection{Beam Steering, Tracking, and Pointing Robustness}

Fig.~\ref{fig:system_level_comparison}(b) shows the comparison between representative beam-steering ranges and the normalized power-consumption indicator. As in Fig.~\ref{fig:system_level_comparison}(a), the horizontal axis represents a qualitative indication of relative power demand. Passive RIS offers wide angular control with low RF power consumption, which looks attractive for coverage shaping and tolerance to pointing variations in moving vehicular D2S terminals.

Phased arrays can provide wide-angle steering, but their main advantage lies in mature and fast electronic beam tracking. Precise beam updates when the terminal or satellite geometry changes are made possible by independent RF control, but at a significantly greater power and hardware expense. Active RIS covers a various design region: its main benefit is the combination of flexible beam control and signal enhancement rather than increasing the steering range alone.

For vehicular D2S links, beamsteering range should also be considered together with beamwidth, tracking speed, control latency, and pointing accuracy. Passive RIS is feasible for energy-efficient coverage control, while phased arrays are more robust for rapid and precise tracking. Active RIS provides an intermediate option when additional link enhancement is required, while hybrid RIS--array architectures may additionally balance steering capability, tracking performance, and terminal power consumption.

\subsection{Scalability and Hardware Complexity}

Fig.~\ref{fig:system_level_comparison}(c) shows the relation between aperture scalability and hardware complexity using normalized qualitative scores. The scores are ordinal indicators meant to show broad architectural trends rather than precise hardware metrics. Passive RIS covers the high-scalability and low-complexity region because enlarging the aperture mainly involves adding controllable unit cells rather than duplicating complete RF transceiver chains.

Phased-array scalability is strongly impacted by the beamforming architecture. The number of RF chains can be reduced by the analog arrays, but hardware complexity increases as more antenna elements, phase shifters, amplifiers, converters, and processing channels are added. This challenge becomes more difficult for highly parallel hybrid and fully digital arrays. Active RIS lies between the two architectures, maintaining metasurface-based aperture scalability while introducing additional biasing, amplification, thermal-management, and reliability requirements.

From a vehicular D2S perspective, passive RIS is more effective when large aperture, low control power, and limited hardware complexity are top concerns. Active RIS offers a balance between aperture scalability and enhanced link performance, whereas phased arrays are preferred when rapid beam tracking, independent RF control, and advanced interference management are essential. The ideal architecture therefore depends on how the terminal balances aperture size, hardware burden, and beamforming functionality.

\subsection{Overall Engineering Interpretation}

Taken together, the three design maps highlight a consistent trade-off among link improvement, power demand, steering capability, aperture scalability, and hardware complexity. The normalized indicators in Fig.~\ref{fig:system_level_comparison} position passive RIS, active RIS, and phased arrays within a common qualitative design space instead of using a direct quantitative comparison among different prototypes.

Phased arrays remain well suited to high-performance vehicular satellite terminals requiring high beamforming gain, rapid electronic tracking, and precise beam control, but these benefits come with higher power demand and hardware complexity. Passive RIS provides low-power operation and favorable aperture scalability, making it more effective for coverage shaping and SWaP-constrained terminals, although the lack of active amplification limits its ability to strengthen an already weak satellite signal. Active RIS partially bridges this gap by combining wavefront control with signal enhancement, at the cost of more power, noise, biasing, and thermal constraints.

The three technologies should therefore not be seen simply as direct competitors. Passive RIS is most effective when power efficiency and aperture scalability dominate, phased arrays is preferred when fast and precise beam tracking is essential, and active RIS offers an intermediate solution when more link enhancement is needed. For future vehicular D2S terminals, hybrid RIS--array architectures may provide a practical balance among tracking reliability, link improvement, aperture scalability, and hardware complexity..

\section{Opportunities, Challenges, and Future Research Directions}
\label{sec:opportunities_challenges}

The comparison of RIS architectures and phased arrays shows that future vehicular D2S communication in satellite networks is unlikely to rely on a single universal architecture. In RIS architectures, passive RIS provides a low-power and energy-efficient solution, but it cannot perform amplification of the signal, which restricts link budget improvement, especially for satellite uplink communication. Active RIS, on the other hand, can improve received signal strength at the cost of fewer RF-chain complexities, but with additional power consumption, noise, and thermal constraints. In electronically steered phased arrays, analog arrays offer efficient directive tracking, while digital and hybrid phased arrays provide stronger multi-beam flexibility at higher RF-chain and processing cost.

\subsection{Opportunities}
Using RIS as a transceiver enables lightweight, low-profile, and energy-efficient vehicular D2S terminals in satellite networks for cars, buses, and trains on the ground, as well as for ships at sea. Passive RIS is attractive for large-aperture coverage shaping with very low DC power consumption, while active RIS can enhance the link budget when passive wavefront control is insufficient. Another promising opportunity is a hybrid RIS-array design, where a compact active feed or small phased-array module provides signal generation and coarse beam steering, while a larger passive or semi-active RIS aperture provides additional wavefront shaping and coverage extension. RIS can also support adaptive coverage and interference management by steering energy toward the intended satellite, reducing leakage toward unintended directions, and assisting spectrum sharing in integrated terrestrial and non-terrestrial networks. With artificial intelligence (AI)-enabled beam management, RIS configurations can be adapted using vehicle location, trajectory, satellite ephemeris, blockage maps, and link-quality history, thereby reducing beam-training overhead and improving handover reliability.

\subsection{Challenges}

Several challenges must be addressed before RIS-based vehicular D2S terminals can be deployed. First, channel estimation and beam training are difficult because passive RIS elements do not have full RF chains, making the cascaded feed--RIS--satellite channel hard to estimate under vehicle mobility and satellite motion. Second, passive RIS suffers from link-budget limitations because it cannot amplify the incident signal; hence, it may be more suitable for downlink enhancement, coverage shaping, or short-to-medium range links unless combined with active or hybrid architectures. Third, active RIS introduces additional power consumption, amplifier noise, thermal load, biasing circuitry, and reliability concerns, especially as the number of active elements increases. Fourth, transmissive RIS terminals must satisfy profile and aerodynamic constraints. A feed-illuminated transmissive RIS can behave like a reconfigurable transmit array, and the feed-to-aperture distance may increase the terminal depth if compact or folded illumination is not used. A realistic deployment of various transmissive RIS architectures requires robust hardware capable of withstanding vibration, rain, dust, temperature fluctuations, radiation, and long-term material degradation. They also need to support low-latency control signaling and integration with satellite modems, tracking systems, and network schedulers.

\subsection{Future Research Directions}

Future research should pay greater attention to hybrid passive--active and RIS--array architectures. These hybrid architectures can provide reliable beam tracking and maintain a basic link budget. Moreover, a larger passive or semi-active aperture can improve coverage shaping and the system's energy efficiency. The design and practical deployment of these architectures will require a joint design of active-element placement, feed configuration, aperture control, thermal management, and terminal profile, and should not be treated as a separate optimization of the communication problem and antenna design.

AI-enabled beam management is another promising research direction that need to be investigated. Vehicle trajectory, inertial sensor data, blockage information, satellite ephemeris, weather conditions, and historical link measurements can be combined by predictive beam controllers to reduce the need for exhaustive beam search and enable faster handovers among different satellites. However, uncertainty-aware protection mechanisms should be incorporated into these learning-based methods to prevent long service interruptions caused by inaccurate predictions.

The idea of RIS-based D2S communication can also be investigated within integrated terrestrial--non-terrestrial networks. RIS can be used for coordinated access, routing, handover, and spectrum management across terrestrial base stations, aerial platforms, and satellite constellations in order to improve coverage, suppress interference, and prepare links for handover. In the case of a large satellite constellation, it is also important to distinguish RIS-generated multi-lobe radiation patterns from independently modulated digital beams. Therefore, it is crucial to carefully evaluate the practical limitations associated with control overhead, synchronization, scheduling, and simultaneous satellite access.

Integrated sensing and communications (ISAC) emerged as a new concept in 6G terrestrial and non-terrestrial networks and offers significant potential. A vehicular D2S terminal can reuse the same aperture for satellite acquisition, blockage detection, localization, road-environment sensing, and data transmission. This will require waveform and aperture designs that jointly balance sensing accuracy, communication rate, beam-tracking delay, and power consumption while satisfying automotive reliability and safety requirements.

Last but not least, meaningful comparisons will require fair link-budget analyses and realistic field trials. Different architectures must be evaluated under common assumptions regarding satellite orbit, carrier frequency, aperture area, EIRP, G/T, bandwidth, receiver noise figure, pointing loss, terminal depth, and implementation losses. Prototype testing will be essential under realistic vehicle mobility, vibration, rainfall, dust, thermal cycling, and control latency to determine the conditions under which passive/ or active RIS, as well as the hybrid RIS--array terminals, can genuinely complement conventional phased arrays.

\section{Conclusion}
\label{sec:conclusion}
The study showed that passive RIS can provide a low-power, energy-efficient solution, but it cannot amplify the signal, which limits link budget improvement, especially for satellite uplink communication. Playing a middle role between passive RIS and phase array, an active RIS can improve received signal strength at the cost of fewer RF-chain complexities, but with additional power consumption, noise, and thermal constraints. In contrast to passive and active RIS, analog arrays offer efficient directive tracking, while digital and hybrid phased arrays provide stronger multi-beam flexibility at higher RF-chain and processing cost. The study analyzed that hybrid RIS--array terminals have the potential to offer the most practical solution for future D2S communication in satellite networks. It can combine the reliability of phased arrays with the energy efficiency and scalability of RIS apertures. The study also highlighted potential opportunities, research challenges, and future directions.

\bibliographystyle{IEEEtran}
\bibliography{Wali_EE}
\end{document}